\documentclass[a4paper]{jpconf}
\usepackage{graphicx}
\begin{document}
\title{Phenomenology of the little bang}

\author{Jean-Yves Ollitrault}

\address{Institut de physique th\'eorique, CNRS/URA2306, CEA Saclay, 91191 Gif-sur-Yvette, France}

\ead{jean-yves.ollitrault@cea.fr}

\begin{abstract}
I review recent selected developments in the theory and modeling of ultrarelativistic heavy-ion collisions. I explain why relativistic viscous hydrodynamics is now used to model the expansion of the matter formed in these collisions. I give examples of first quantitative predictions, and I discuss remaining open questions associated with the description of the freeze-out process. 
I argue that while the expansion process is now well understood, our knowledge of initial conditions is still poor. 
Recent analyses of two-particle correlations have revealed fine structures known as ridge and shoulder, which extend over a long range in rapidity. These correlations are thought to originate from initial state fluctuations, whose modeling is still crude.
I discuss triangular flow, a simple mechanism recently put forward, through which fluctuations generate the observed correlation pattern.  
\end{abstract}    

\section{Introduction}
The Relativistic Heavy Ion Collider (RHIC) at Brookhaven has been running for ten years, and the Large Hadron Collider (LHC) at CERN is expected to collide nuclei before the end of 2010. In the three years since the last INPC conference, we have witnessed significant progress in our understanding of RHIC experiments, both at the quantitative and qualitative level. 
A recent breakthrough is that quantitative predictions from viscous relativistic hydrodynamics are now
available~\cite{Luzum:2008cw,Luzum:2009sb}, which allow for a better description of the bulk of observables. 
Section~\ref{s:hydro} focuses on this topic, although there have been important developments in ideal hydrodynamics 
as well \cite{Broniowski:2008vp,Andrade:2008xh,Huovinen:2009yb}.  
On the experimental side, new analyses involve rare particles, such as heavy flavours, and detailed investigations of the fine structure of correlations. I concentrate on the latter topic in section~\ref{s:correlations}, and I discuss a new mechanism which sheds new light on the ``ridge'' and ``shoulder'' structures observed in correlations, which have puzzled theorists for a few years. 

\section{Relativistic viscous hydrodynamics}
\label{s:hydro}

Nuclei colliding at RHIC are thin pancakes: the Lorentz contraction factor is 100 at the top energy. 
The clear separation between the small longitudinal size and the much larger transverse size $R$ is a 
key feature of heavy-ion phenomenology. In a time short compared with $R$ (in natural units $c=1$), matter 
is produced: a system of strongly interacting quarks and gluons, which rapidly expands into the vacuum. 
The underlying microscopic description is a quantum field theory, quantum chromodynamics, in the nonperturbative regime, which 
does not necessarily admit a kinetic description in terms of quasiparticles. However, if the system is large
enough, hydrodynamics gives a reliable description of its real-time macroscopic evolution, once 
{\it initial conditions\/} are specified, at a time $t\ll R$. 

For a nonrelativistic fluid, the usual form of the Navier-Stokes equation is
\begin{equation}
\rho\left(\frac{\partial\vec v}{\partial t}+\vec v\cdot\vec\nabla\vec v\right)=-\vec\nabla P+\eta\nabla^2\vec v.
\label{navier}
\end{equation}
where $\rho$ is the mass density, $\vec v$ the fluid velocity, $P$  the local pressure, and $\eta$ the shear viscosity. 
Let us analyze the order of magnitude of the various terms. Our analysis will differ from that usually  
found in textbooks. Usually, the Navier-Stokes equation is used in the
context of {\it incompressible\/} flows. 
Incompressible does not refer to an intrinsic property of
matter. It is a synonym of   
slow: it characterizes a flow where $v$ is much smaller than
microscopic (thermal) velocities $v_{\rm th}$.  
In this case (which applies for example to the lower atmosphere), 
it can be shown that the 
density is almost constant throughout the fluid, hence the term
``incompressible''.  
In a heavy-ion collision, however, the fluid is freely expanding into
the vacuum, which is a clear instance of a compressible flow.  
Transverse expansion is driven by the thermal motion, so that $v\sim v_{\rm th}$. 
Gradients and time derivatives all scale like
 $1/R$. In Eq.~(\ref{navier}), all terms scale like $1/R$, except the
 viscous term which scales like $1/R^2$.  
Its magnitude relative to the acceleration term is $\lambda/R$, where
$\lambda$ is a length involving thermodynamic quantities. 
Inspection of  equation~(\ref{navier}) gives $\lambda=\eta/\rho v_{\rm th}$. 
In systems which can be described by kinetic theory, $\lambda$ is the particle mean free path. 
Now, fluid dynamics is a macroscopic description which is valid only if $\lambda\ll R$. 
Therefore the viscous term must be a small correction, or fluid dynamics itself breaks down. 

The dynamics of compressible fluids can be formulated as a systematic gradient expansion, that is, an expansion in the Knudsen number $K\equiv \lambda/R$~\cite{Baier:2007ix}. 
Ideal hydrodynamics is the leading order, and Navier-Stokes theory is the first correction of order $K$. 
The second-order correction, of order $K^2$, is usually called Israel-Stewart theory in the relativistic case. 
The expansion converges if $K$ is small: hydrodynamics applies to heavy-ion collisions only if the 
nucleus is large enough, and if the viscosity $\eta$ is small enough. 

Until 2007, the state of the art for RHIC phenomenology was ideal hydrodynamics~\cite{Kolb:2003dz}, 
motivated by the hope that the viscosity of hot QCD is very small~\cite{Policastro:2001yc,Kovtun:2004de}.
It was realized early on that viscous corrections were likely to be important, both on 
theoretical grounds \cite{Teaney:2003kp} and from a critical analysis of 
RHIC data~\cite{Bhalerao:2005mm}.
However, going beyond ideal hydrodynamics proved a difficult task. 
Relativistic Navier-Stokes equations are known to break causality~\cite{Hiscock:1985zz} because
small-wavelength modes propagate faster than light. This is not a conceptual 
\begin{figure}[h]
\includegraphics[width=25pc]{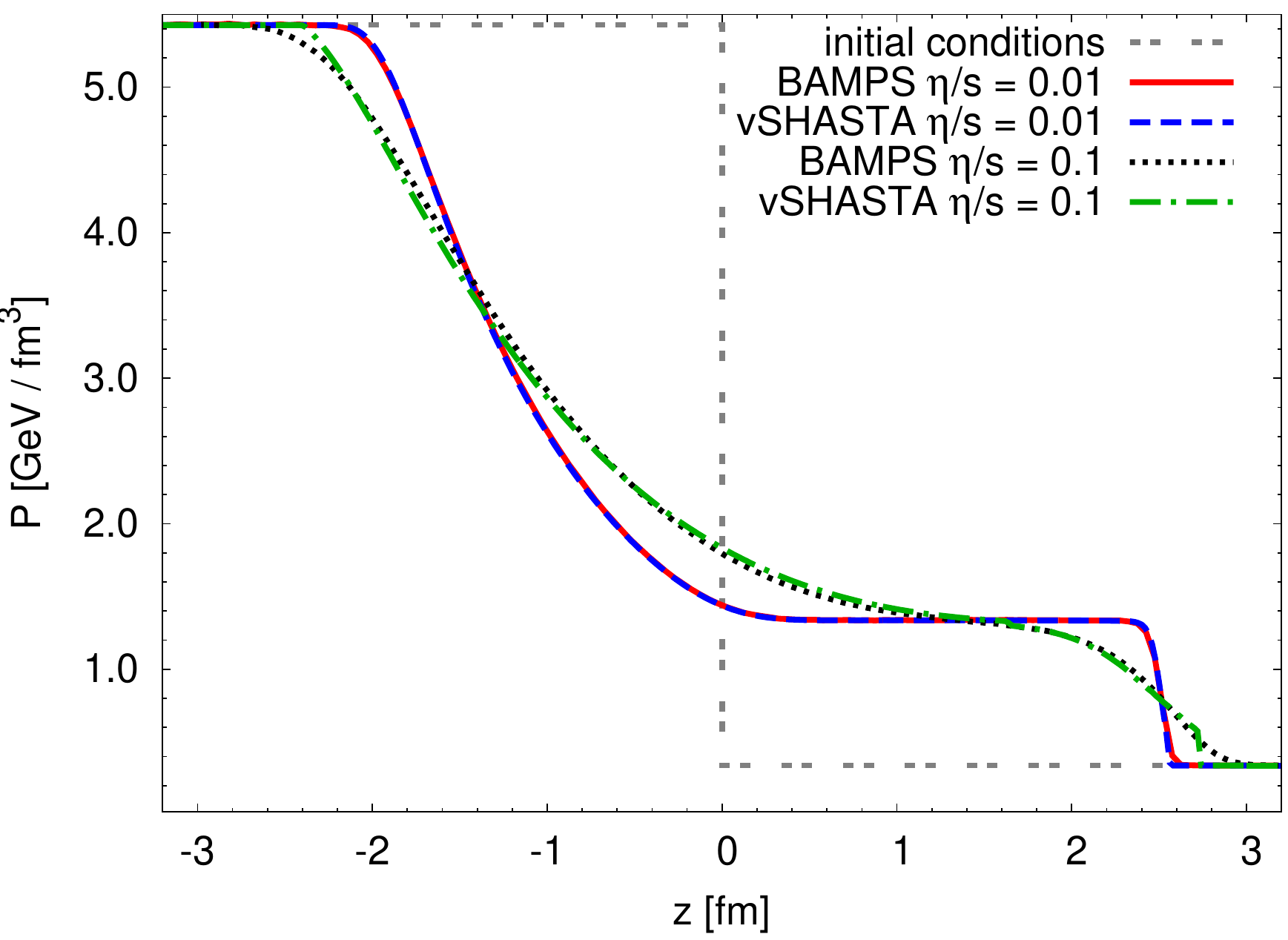}\hspace{2pc}%
\begin{minipage}[b]{11pc}\caption{\label{ioannis}(courtesy I. Bouras \cite{Bouras:2010hm}) Comparison between second-order viscous hydrodynamics (vSHASTA) and kinetic theory (BAMPS), for two values of the viscosity to entropy ratio $\eta/s$. The dashed line is the initial condition at $t=0$, and the curves display the pressure profile at $t=3.2$~fm/c.}
\end{minipage}
\end{figure}
problem, since hydrodynamics is only expected to describe long-wavelength modes~\cite{Geroch:1995bx}, 
but it is a practical difficulty because of the associated numerical instabilities. 
As a consequence, one must include second-order terms. 
There were early attempts to apply second-order viscous hydrodynamics to heavy-ion 
collisions~\cite{Muronga:2001zk}, but full calculations, including comparison 
with RHIC data, became available only 
recently~\cite{Romatschke:2007mq,Song:2007fn,Denicol:2009am,Monnai:2009ad,Bozek:2009dw}.
There is some arbitrariness in second-order terms, but the goal is to estimate 
quantitatively the first-order correction, in regimes where second-order terms are negligible. 

If second-order terms could be determined from first principles, they would extend the 
validity of the hydrodynamic description up to larger values of the Knudsen number. 
Although this is not yet the case for QCD, a feasibility study was recently 
performed~\cite{Bouras:2010hm} for a dilute system described by kinetic theory, 
where second-order terms were determined unambiguously. 
Figure~\ref{ioannis}  displays
the comparison between hydrodynamics (labeled vSHASTA) 
and kinetic theory (labeled BAMPS). 
At time $t=0$, the fluid is at rest, 
and its pressure is uniform with a discontinuity at $z=0$ (shock tube problem). 
At $t>0$, the fluid flows toward lower pressures, i.e., to the right. 
In ideal hydrodynamics, the flow profile consists of three distinct parts: a continuous 
rarefaction wave propagating to the left, a compression shock (pressure discontinuity)
propagating to the right, and a region of uniform flow in between.
This pattern is clearly seen with the smaller value of the viscosity $\eta/s=0.01$. 
If the viscosity is increased by a factor of 10, the flow pattern is smoothed. In particular, 
the width of the shock is proportional to the viscosity. 
In both cases, there is almost perfect agreement between hydrodynamics and kinetic
theory. Agreement is found to hold within 10\%  
up to $K=1/2$. Both the time scale in figure~\ref{ioannis} and the larger value  
$\eta/s=0.1$ are of the order of expected values at RHIC, so that the difference between 
$\eta/s=0.01$ and $\eta/s=0.1$ results gives an idea of the magnitude of viscous effects at RHIC.

\begin{figure}[h]
\begin{minipage}{18pc}
\includegraphics[width=18pc]{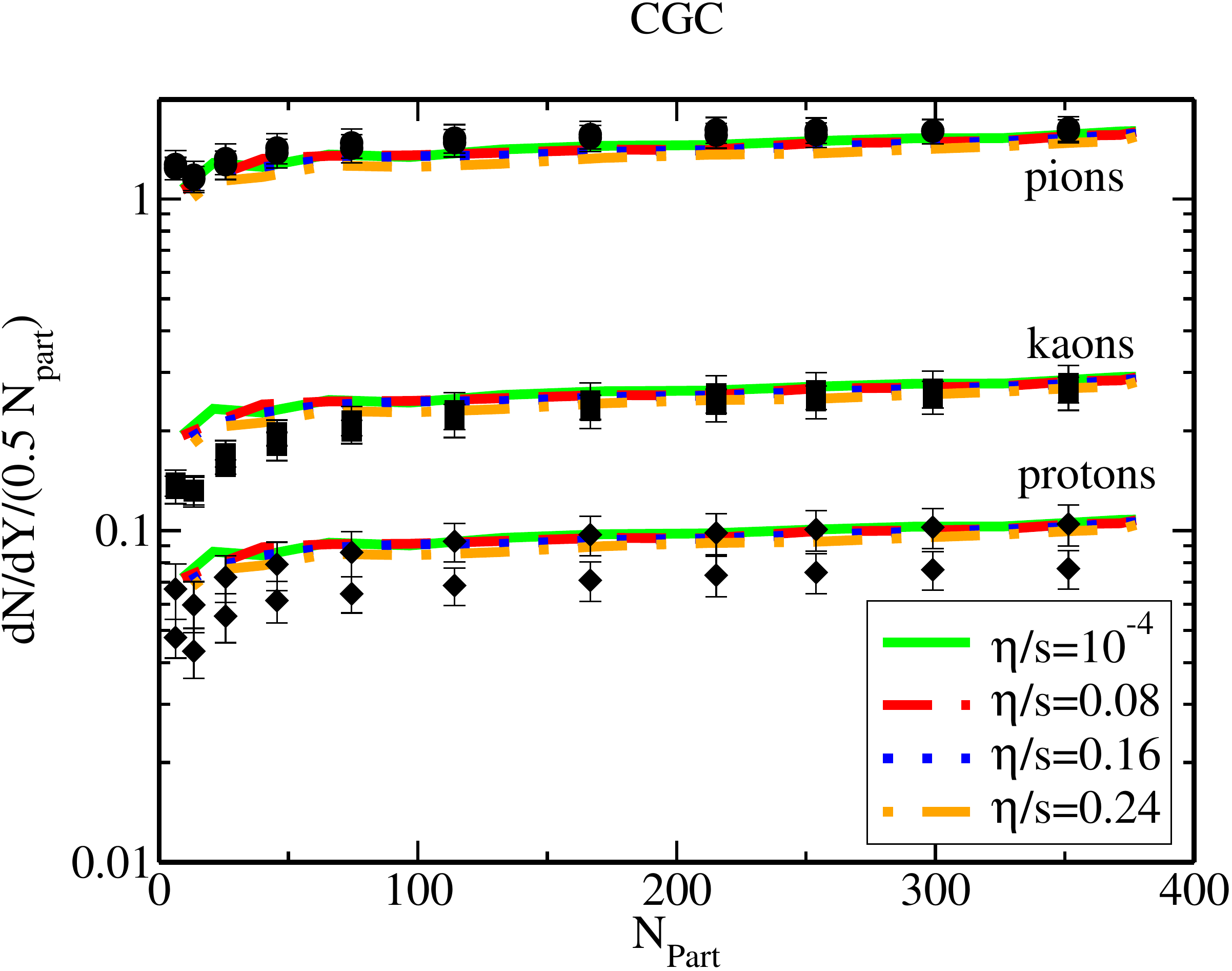}
\caption{\label{mattdndy}
Average multiplicity of pions, kaons, protons, and antiprotons 
 per participant, versus number of participants (from \cite{Luzum:2008cw}).
The number of participant nucleons is used to estimate the centrality
of the collision. It is maximum for central, head-on collisions.
\phantom{coucou} 
\phantom{coucou} 
\phantom{coucou} 
\phantom{coucou} 
}
\end{minipage}\hspace{2pc}%
\begin{minipage}{18pc}
\includegraphics[width=18pc]{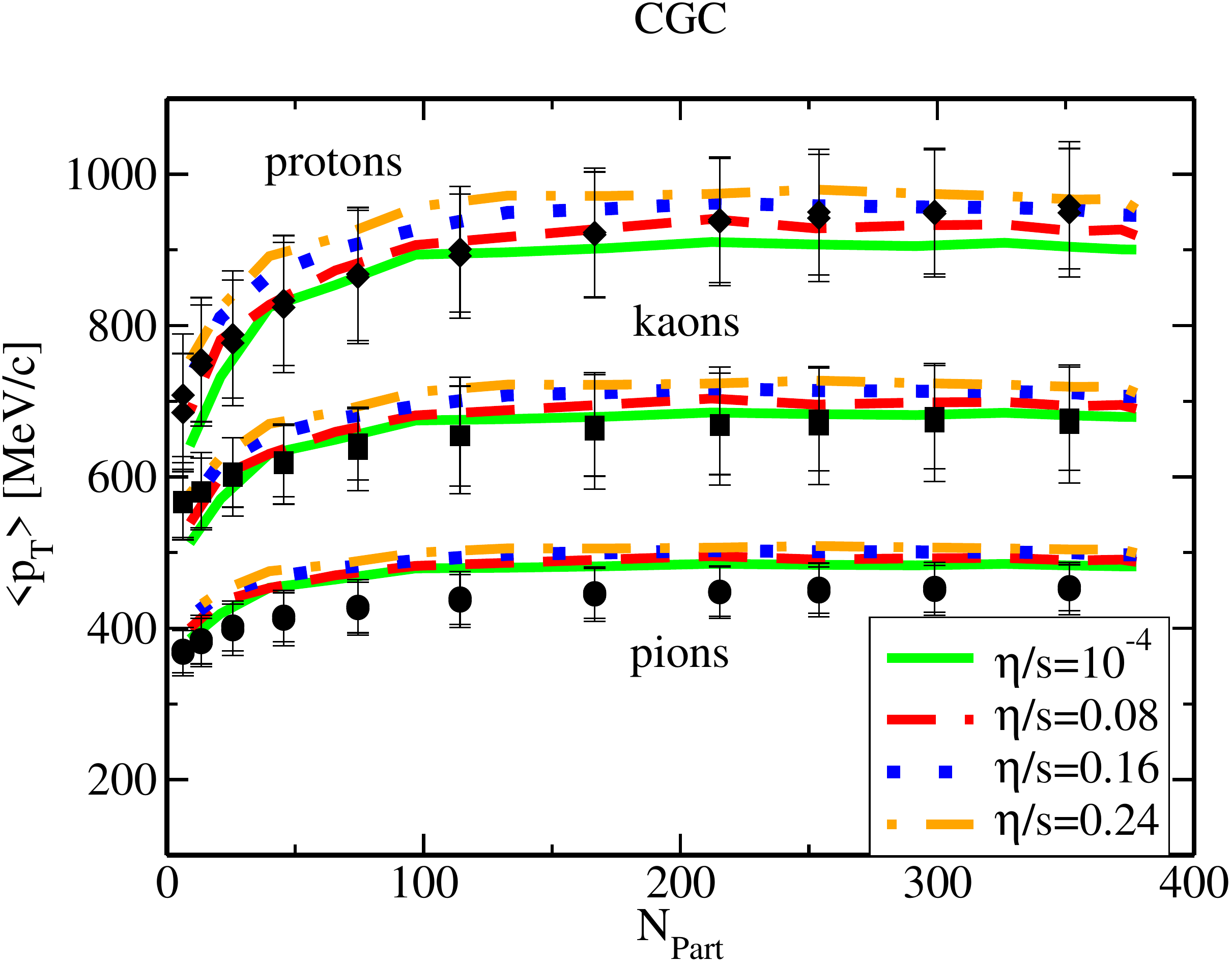}
\caption{\label{mattpt}
Average transverse momentum of particles versus number of participants (from \cite{Luzum:2008cw}). As in figure~\ref{mattdndy}, experimental data from PHENIX~\cite{Adler:2003cb} are compared with viscous hydrodynamics calculations. The label CGC refers to a specific choice of the initial density profile in hydrodynamics.}
\end{minipage} 
\end{figure}
I now discuss hydrodynamic calculations for collisions at RHIC. The initial energy profile in the transverse plane cannot be determined from first principles because interactions are largely nonperturbative. This is a major source of uncertainty.
Most calculations use a simple ansatz which reproduces the observed centrality dependence of the pion multiplicity (figure~\ref{mattdndy}). The multiplicity scales essentially like the number of nucleons participating in the collision. Particle ratios (proton to pion and kaon to pion) are also essentially constant. In hydrodynamics, they are determined by the freeze-out temperature at which the fluid transforms into hadrons. 
The average transverse momentum $\langle p_t\rangle$ of identified particles  (figure~\ref{mattpt}) 
also shows little centrality dependence.  
This observable reflects the kinetic energy per particle. In
hydrodynamics, the kinetic energy is 
proportional to the effective temperature of the system, defined as
the average temperature at a time of order $R$ when transverse
expansion fully develops.  
This temperature is determined (through the equation of state) by 
the density of particles per unit volume, which is itself
closely related to the multiplicity per participant plotted in
figure~\ref{mattdndy}. Since this density depends little on
centrality, hydrodynamics naturally predicts that $\langle p_t\rangle$ 
depends little on centrality~\cite{Ollitrault:1991xx}.  
While the absolute scale of  $\langle p_t\rangle$  depends little on
viscosity, it depends crucially on the equation of state used in the
hydrodynamic calculation, which is taken from lattice QCD
calculations. The equation of state determines the 
temperature as a function of the density, and the fact that
hydrodynamics reproduces data is a non trivial test of the equation of
state from lattice QCD.  

\begin{figure}[h]
\includegraphics[width=24pc]{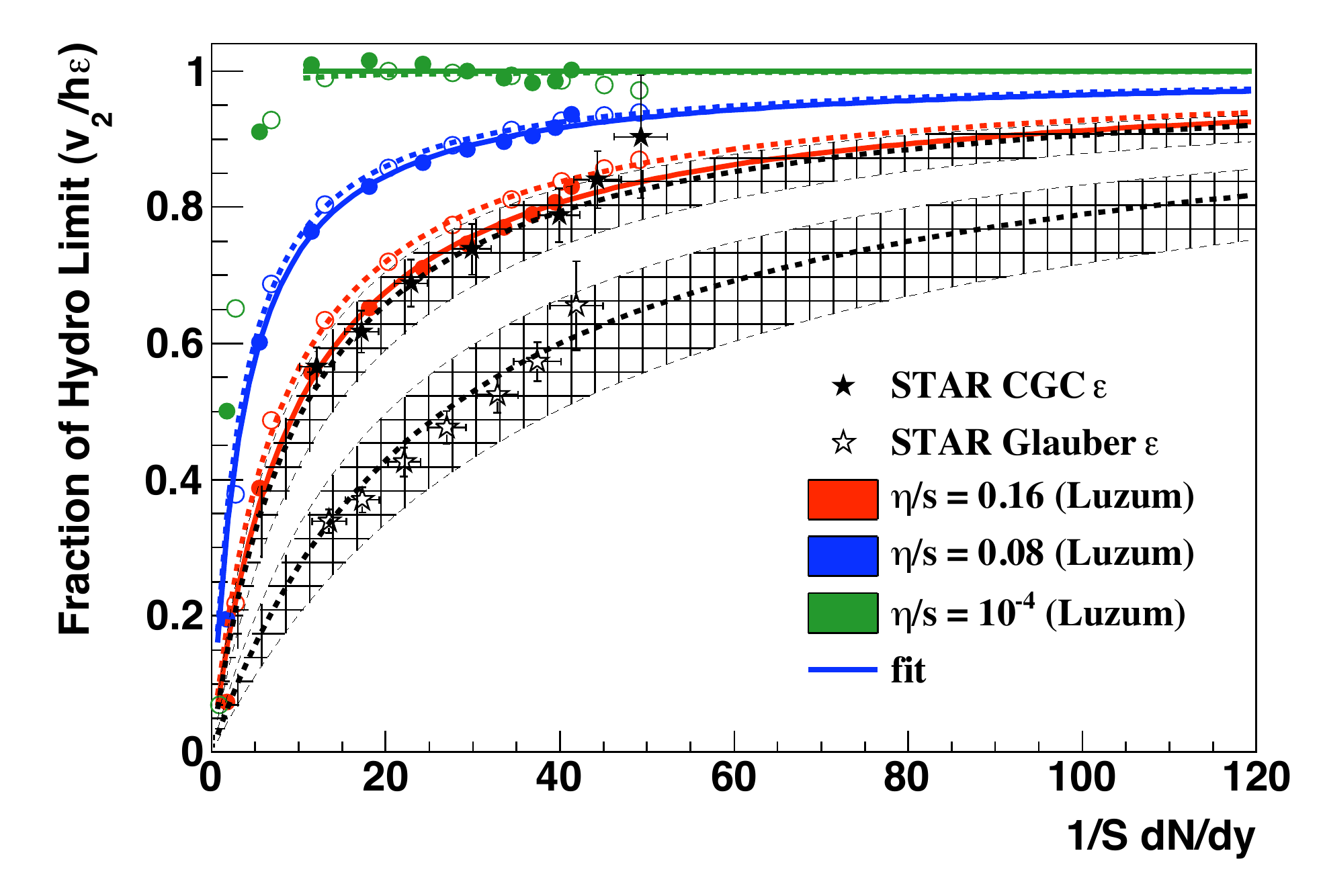}
\begin{minipage}[b]{14pc}\caption{\label{raimond} Elliptic flow scaled
    by the initial eccentricity $\varepsilon$ and by the hydrodynamic
    limit ${\rm h}$ (see text) 
as a function of the density of particles per unit area. Results from ideal and viscous hydrodynamics~\cite{Luzum:2008cw} are plotted together with data from 
RHIC (from \cite{Masui:2009pw}). Open symbols and full symbols correspond to two different choices of initial conditions (called CGC and Glauber). Dashed and solid lines are fits using equation~(\ref{fit}).}
\end{minipage}
\end{figure}
I now discuss an observable which is more sensitive to viscosity, namely, elliptic flow. The azimuthal distribution of emitted particles is elliptic, due to the elliptic shape of the interaction region in the transverse plane (see figure~\ref{burakellipse}). 
The fluid velocity is proportional to the pressure gradient, which is larger along the smaller dimension.
Particles tend to be emitted parallel to the fluid velocity, which results in more particles emitted along the minor 
axis~\cite{Ollitrault:1992bk}. The corresponding observable is  
\begin{equation}
v_2\equiv \langle \cos(2\phi-2\Psi_2)\rangle,
\label{defv2}
\end{equation}
where $\Psi_2$ is the minor axis of the participant ellipse in
figure~\ref{burakellipse}, $\phi$ is the azimuthal angle of an
outgoing particle, and angular brackets denote an average over
particles and events. $v_2$ is smallest for central collisions, where
the overlap area is circular except for fluctuations, and increases up
to 7\%   for more peripheral collisions. Since elliptic flow is
generated by pressure, it is a clean signature of collective flow.  

Figure~\ref{raimond} displays the centrality dependence of elliptic
flow, plotted in such a way that results can be easily understood by
dimensional arguments. The vertical axis displays $v_2$ scaled by the
initial eccentricity, $\varepsilon$, of the participant ellipse in
figure~\ref{burakellipse}. The horizontal axis is the multiplicity of
produced particles  scaled by the area of the ellipse.  
This density per unit area is often used as an alternative measure of
the centrality. Since the effective density of particles per unit
volume depends little on centrality, $(1/S)(dN/dy)$ is proportional to 
the system size $R$: the largest value corresponds to central 
collisions, as in figures~\ref{mattdndy} and \ref{mattpt}. 
In hydrodynamics, $v_2/\varepsilon$ depends little on the details of
initial conditions. It is essentially determined by the size $R$ and
by thermodynamic quantities, such as the speed of sound and the 
viscosity. 
Thermodynamic quantities depend on the temperature, but we 
have seen that the effective temperature depends little on centrality, 
so that $v_2/\varepsilon$ depends on centrality only through the size $R$. 
In ideal hydrodynamics (top line in figure~\ref{raimond}), 
$v_2/\varepsilon$ is independent of the system size. 
This is a natural 
consequence of the scale invariance of equation~(\ref{navier}) with 
$\eta=0$: velocity and pressure are unchanged under the transformation 
$(t,\vec x)\to  (\alpha t,\alpha\vec x)$, where $\alpha$ is 
arbitrary. 
With a nonzero viscosity, thermalization is incomplete. Therefore, 
there is less pressure and less flow, and $v_2/\varepsilon$ is 
smaller. Using the dimensional analysis explained at the beginning of 
this section, the viscous correction is expected to scale with $\eta$ 
and $R$ like $\eta/R\propto \eta ((1/S)(dN/dy))^{-1}$. A look at the 
viscous hydro results for $\eta/s=0.08$ and $\eta/s=0.16$ in
figure~\ref{raimond} shows that the scaling is at least qualitatively
correct: deviations from ideal hydrodynamics are proportional to the
viscosity, and inversely proportional to $(1/S)(dN/dy)$. 
In order to check the scaling quantitatively, each set of results
(given value of $\eta/s$ and choice of initial conditions) is  fit
using the two parameter formula 
\begin{equation}
\frac{v_2}{\varepsilon}=\frac{{\rm h}}{1+B\left(\frac{1}{S}\frac{dN}{dy}\right)^{-1}}.
\label{fit}
\end{equation}
The parameter ${\rm h}$ is the ideal hydro limit. The vertical axis in
figure~\ref{raimond} is scaled by ${\rm h}$ in order to single out the viscous correction, quantified by the parameter $B$. This parameter is found to be proportional to $\eta/s$, as expected. Unlike $\langle p_t\rangle$, elliptic flow is very sensitive to the viscosity. 
$\eta/s=0.08$ is close to an absolute lower
bound~\cite{Kovtun:2004de}, yet it produces sizable effects, because
an atomic nucleus is not quite a macroscopic object. Note that both
sets of initial conditions (CGC and Glauber) give almost the same
$v_2/\varepsilon$, although $v_2$ and $\varepsilon$ differ.  

The next step is to put experimental data on this plot. $v_2$ and $dN/dy$ are measured, but $\varepsilon$ and $S$ must be calculated within a model, which entails uncertainties~\cite{Nagle:2009ip}. 
Two different models of initial conditions are used, and the results are then fit using equation~(\ref{fit}). Both models take into account eccentricity 
fluctuations~\cite{Alver:2006wh} due to the finite number of nucleons (see figure~\ref{burakellipse} and section~\ref{s:correlations}), which are important for central collisions. 
Equation~(\ref{fit}) fits experimental data quite well~\cite{Drescher:2007cd}.
By comparing theory with data, we can extract the viscosity of QCD from this plot. Depending on which set of initial conditions is used, the value of $\eta/s$ is 0.16 or significantly larger. Our poor knowledge of the initial density profile hinders our ability to extract thermodynamic quantities from the data. 
With CGC initial conditions, the viscous correction is roughly 20\%  
for central collisions. It is small enough for hydrodynamics to be a valid approach; but it is sizable, and viscosity must be taken into account in order to achieve quantitative agreement with data, in particular for elliptic flow.

\begin{figure}[h]
\begin{minipage}{18pc}
\includegraphics[width=18pc]{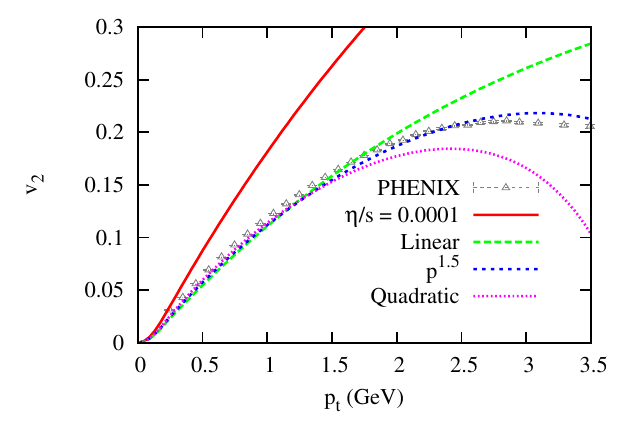}
\caption{\label{mattv2}Elliptic flow versus transverse momentum in mid-central Au-Au collisions at RHIC. Calculations are shown for pions~\cite{Luzum:2010ad} using ideal hydrodynamics ($\eta/s=0.0001$) and viscous hydrodynamics ($\eta/s=0.16$) with several choices of $\delta f_{\rm viscous}$ are shown together with recent data for charged hadrons from the PHENIX collaboration~\cite{Adare:2010ux}.}
\end{minipage}\hspace{2pc}%
\begin{minipage}{18pc}
\includegraphics[width=18pc]{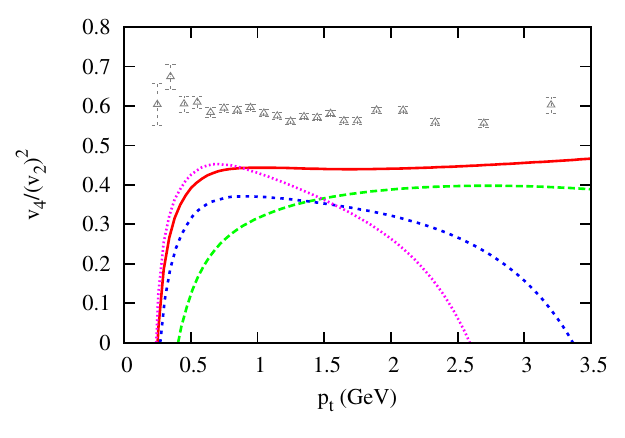}
\caption{\label{mattv4}
The ratio of quadrangular flow, $v_4$, to elliptic flow squared, versus transverse momentum. 
Theory as in figure~\ref{mattv2}. 
PHENIX data have been scaled down by a constant factor to take into
account fluctuations of the initial geometry (eccentricity
fluctuations) which increase $v_4/(v_2)^2$~\cite{Gombeaud:2009ye}. 
\phantom{coucou} 
\phantom{coucou} 
\phantom{coucou} 
}
\end{minipage} 
\end{figure}
I conclude this section by mentioning some of the pending issues in hydrodynamics. 
While viscous hydrodynamics provides us with a detailed picture of the expansion, we 
still miss the beginning and the end of the story: the uncertainty in the initial density profile
has already been pointed out. 
There are also uncertainties in the final state, when the fluid transforms into particles. 
This freeze-out procedure is well defined in ideal hydrodynamics: thermal equilibrium is 
achieved locally, so that momentum distributions are thermal at any point in the fluid rest frame. 
Viscosity leads to a small deviation from thermal equilibrium, and the momentum distribution in the fluid rest frame is
\begin{equation}
\label{deltaf}
\frac{dN}{d^3 x d^3 p}\propto e^{-E/T}\left(1+\delta f_{\rm viscous}(\vec p)\right),
\end{equation}
where I have neglected quantum statistics for simplicity. 
The dependence of $\delta f_{\rm viscous}(\vec p)$ on the direction of $\vec p$ is constrained by the continuity of the pressure tensor at freeze-out, but its dependence on $p\equiv |\vec p|$ is not. Most calculations so far assume 
$\delta f_{\rm viscous}\propto p^2$~\cite{Teaney:2003kp}, but it has been recently emphasized~\cite{Dusling:2009df} that other choices are possible. The correct choice depends on the 
detailed structure of hadronic cross sections at freeze-out. 
Figure~\ref{mattv2} displays a comparison between theory and data for elliptic flow.
While ideal hydrodynamics predicts too large a $v_2$, as expected from the previous discussion, viscous hydrodynamics matches the data. Results up to $p_t\sim 2$~GeV are in fact fairly insensitive to the choice of $\delta f_{\rm viscous}$, which is good news. For momenta larger than 2~GeV, results depend on $\delta f_{\rm viscous}$, but the viscous correction becomes large and hydrodynamics breaks down anyway. 
Now, look at figure~\ref{mattv4}, which displays results for the fourth harmonic of the azimuthal distribution. $v_4$ is defined by 
\begin{equation}
v_4\equiv \langle \cos(4\phi-4\Psi_2)\rangle.
\label{defv4}
\end{equation}
Once scaled by $(v_2)^2$, data are essentially flat as a function of transverse momentum. While ideal hydrodynamics naturally reproduces this flat behaviour, taking into account viscosity makes the agreement with data {\it worse\/}, even for moderate values of $p_t$ where hydrodynamics is expected to be reliable. Among the three choices of $\delta f_{\rm viscous}$, the standard quadratic ansatz, which is used essentially by all viscous hydro calculations so far, gives the worst results. 

Another shortcoming of present viscous hydrodynamics calculations is
that they do not incorporate chemical freeze-out, i.e., the
observation that ratios of particle abundances are constant below some
temperature (typically 165 MeV). In ideal hydrodynamics, entropy is
conserved, so that the entropy per particle is uniform throughout
after chemical freeze-out. This makes chemical freeze-out easy to
implement~\cite{Hirano:2002ds}. In viscous hydrodynamics, diffusion
processes must be taken into account, resulting in complicated
equations~\cite{Monnai:2010qp}. An alternative way of dealing with the
hadronic phase is to couple hydrodynamics to kinetic theory below some
temperature. This was done in the early days of RHIC for ideal 
hydrodynamics~\cite{Bass:2000ib,Teaney:2000cw,Hirano:2005xf,Nonaka:2006yn,Petersen:2008dd}
and is being generalized to viscous hydrodynamics. This approach also
has problems, because the coupling is done at a high density where the
validity of kinetic theory is questionable. 

\section{Correlations and fluctuations}
\label{s:correlations}

\begin{figure}[h]
\begin{minipage}{18pc}
\includegraphics[width=18pc]{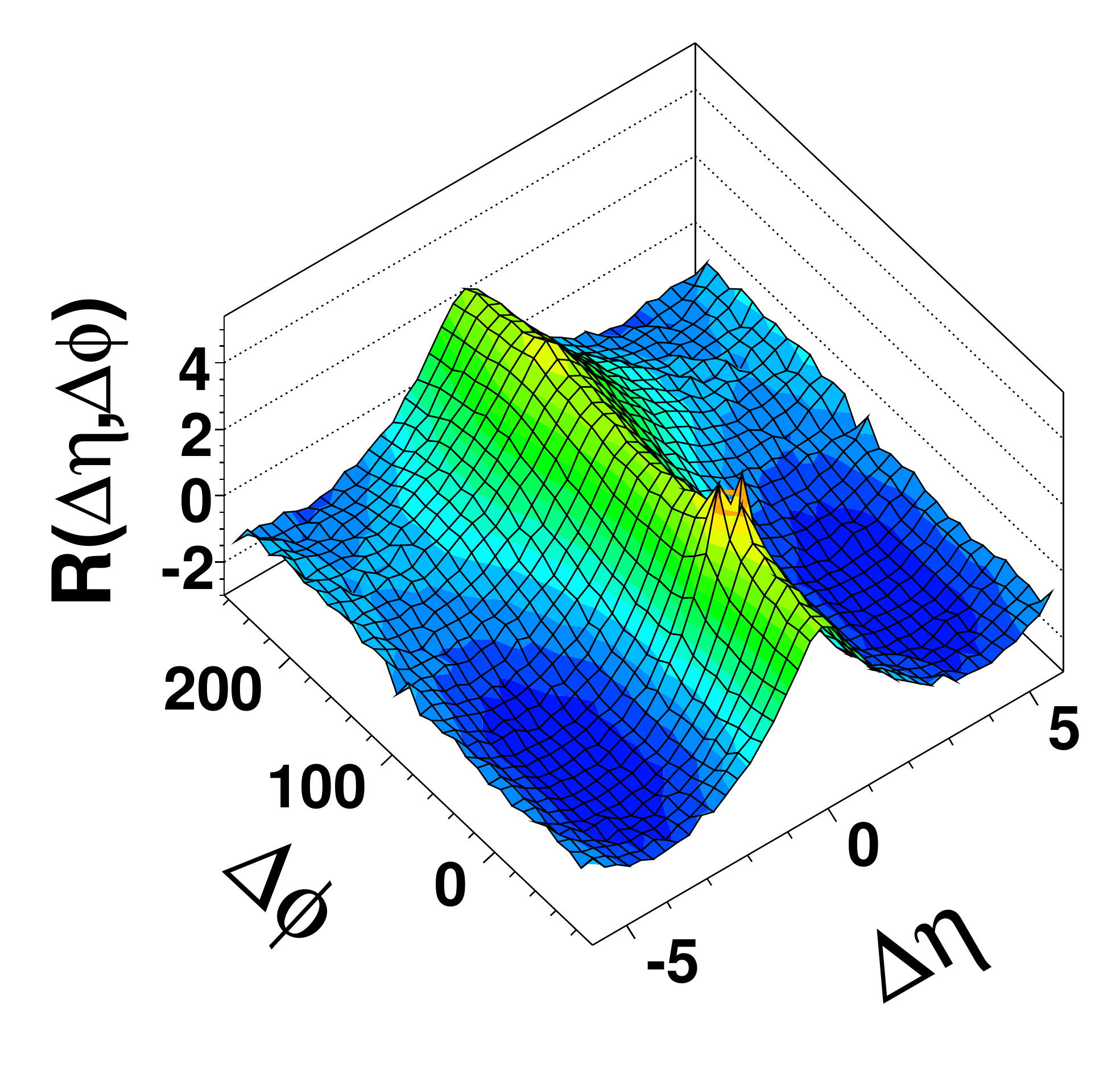}
\caption{\label{phobospp}Measured two-particle correlation in proton-proton collisions at centre-of-mass energy 200~GeV.~\cite{Alver:2008gn}. }
\end{minipage}\hspace{2pc}%
\begin{minipage}{18pc}
\includegraphics[width=18pc]{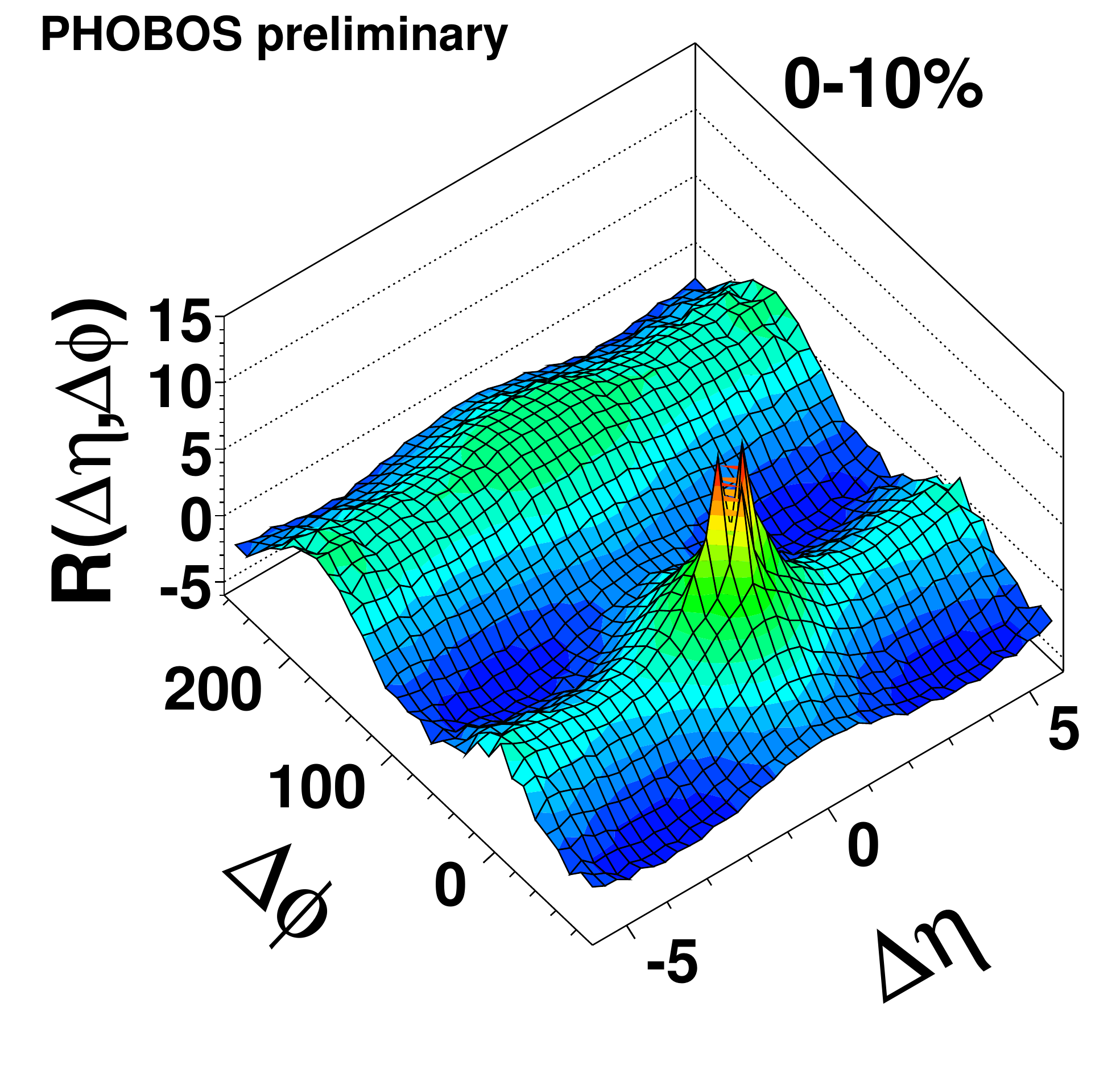}
\caption{\label{phobosAu}Measured two-particle correlation in central Au-Au collisions, same energy per nucleon as in figure~\ref{phobospp}.}
\end{minipage} 
\end{figure}
The most striking differences between proton-proton and nucleus-nucleus collisions are probably found in correlations. 
I first define a measure of the correlation. 
In a given event, let $N_1$ and $N_2$ denote the numbers of particles in two different phase-space bins, say, with azimuthal angles  $\phi_1$ and $\phi_2$, up to some $\delta\phi$. 
The correlation is defined by 
\begin{equation}
{\cal C}\equiv \frac{\langle N_1N_2\rangle}{\langle N_1\rangle\langle N_2\rangle}-1,
\end{equation}
where angular brackets denote an average over a larger number of events within a centrality class. 
Correlations typically vary like the inverse of the system size: if a nucleus-nucleus was a superposition of $n$ independent nucleon-nucleon collisions, ${\cal C}$ would be proportional to $1/n$. In order to remove this trivial size dependence, the PHOBOS collaboration multiplies the correlation by the average number of particles~\cite{Alver:2008gk}. 
The correlation is measured as a function of the relative azimuthal angle 
$\Delta\phi\equiv\phi_1-\phi_2$, and of the relative pseudorapidity $\Delta\eta$. The pseudorapidity is a function of the longitudinal velocity, parallel to the beam.
Results are displayed for proton-proton collisions in figure~\ref{phobospp}, and for Au-Au collisions in  figure~\ref{phobosAu}. The difference is striking. In proton-proton collisions, the correlation is peaked around $\Delta\eta=0$: correlated particles have similar longitudinal velocities. 
This correlation is essentially flat in azimuth, except for a small peak near $\Delta\phi=0$ and a wider peak near $\Delta\phi=\pi$.
In Au-Au collisions, on the contrary, the correlation has a marked structure in $\Delta\phi$, which extends to large $\Delta\eta$.  

Simple causality arguments~\cite{Dumitru:2008wn} show that correlations at large 
 $\Delta\eta$  can only be created at early times. It is in fact likely that their source is contained in the wavefunction of the incoming projectiles. 
The $\Delta\phi$ dependence of the correlation function at large $\Delta\eta$ has two maxima at 
$\Delta\phi=0$ and $\Delta\phi=\pi$, which can be explained by a term proportional to $\cos 2\Delta\phi$. This term is due to elliptic flow. 
As explained above, the origin of elliptic flow lies in the elliptic shape of the overlap area.
Assuming symmetry with respect to the direction of $\psi_2$, one can
rewrite equation (\ref{defv2}) in complex notation, for a single event,  
\begin{equation}
\label{otherdefv2}
\langle e^{2 i \phi}\rangle=v_2 e^{2 i \Psi_2}.
\end{equation}
Assuming that the only correlation is due to flow, 
\begin{equation}
\label{v22}
\langle\cos 2\Delta\phi\rangle=
\langle e^{2 i (\phi_1-\phi_2)}\rangle=\langle e^{2 i \phi_1}\rangle
\langle e^{-2 i\phi_2}\rangle=(v_2)^2.
\end{equation}
This well-known correlation is usually subtracted from the observed correlation. 

\begin{figure}[h]
\includegraphics[width=19pc]{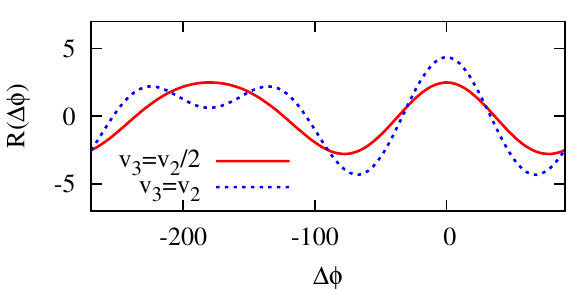}\hspace{2pc}%
\begin{minipage}[b]{17pc}\caption{\label{corr}
Pair azimuthal correlation 
$R(\Delta\phi)\propto
v_2^2\cos 2\Delta\phi+v_3^2\cos 3\Delta\phi-b\cos\Delta\phi$
from elliptic and triangular flow. 
I have chosen $b=\frac{1}{4} v_2^2$ for both curves. Shoulders appear
around $\Delta\phi=\pi$ if $v_3>\frac{2}{3}v_2$.}
\end{minipage}
\end{figure}

A closer scrutiny of figure~\ref{phobosAu} reveals that the
correlation at large $\Delta\eta$ is not exactly a $\cos
2\Delta\phi$ modulation. While the maxima at $\Delta\phi=0$ and 
$\Delta\phi=\pi$ have approximately the same height, the
distribution is narrower around the first maximum at $\Delta\phi=0$. 
The narrower peak around $\Delta\phi=0$ has been dubbed the {\it ridge\/} and has puzzled theorists for several 
years~\cite{Dumitru:2008wn,Gavin:2006xd,Shuryak:2007fu,Gavin:2008ev,Takahashi:2009na}.
One easily checks that this pattern can be reproduced by assuming, in
addition to the $\cos 2\Delta\phi$ term, a term $a\cos
3\Delta\phi-b\cos\Delta\phi$ (see figure~\ref{corr}). The maxima at
$\Delta\phi=0$ and $\Delta\phi=\pi$ have the same height if $a=b$. 
Transverse momentum conservation naturally generates a 
$-\cos\Delta\phi$ term~\cite{Borghini:2000cm}, but there was no known
mechanism to produce a $\cos 3\Delta\phi$ correlation.  

\begin{figure}[h]
\begin{minipage}{18pc}
\begin{center}
\includegraphics[width=13pc]{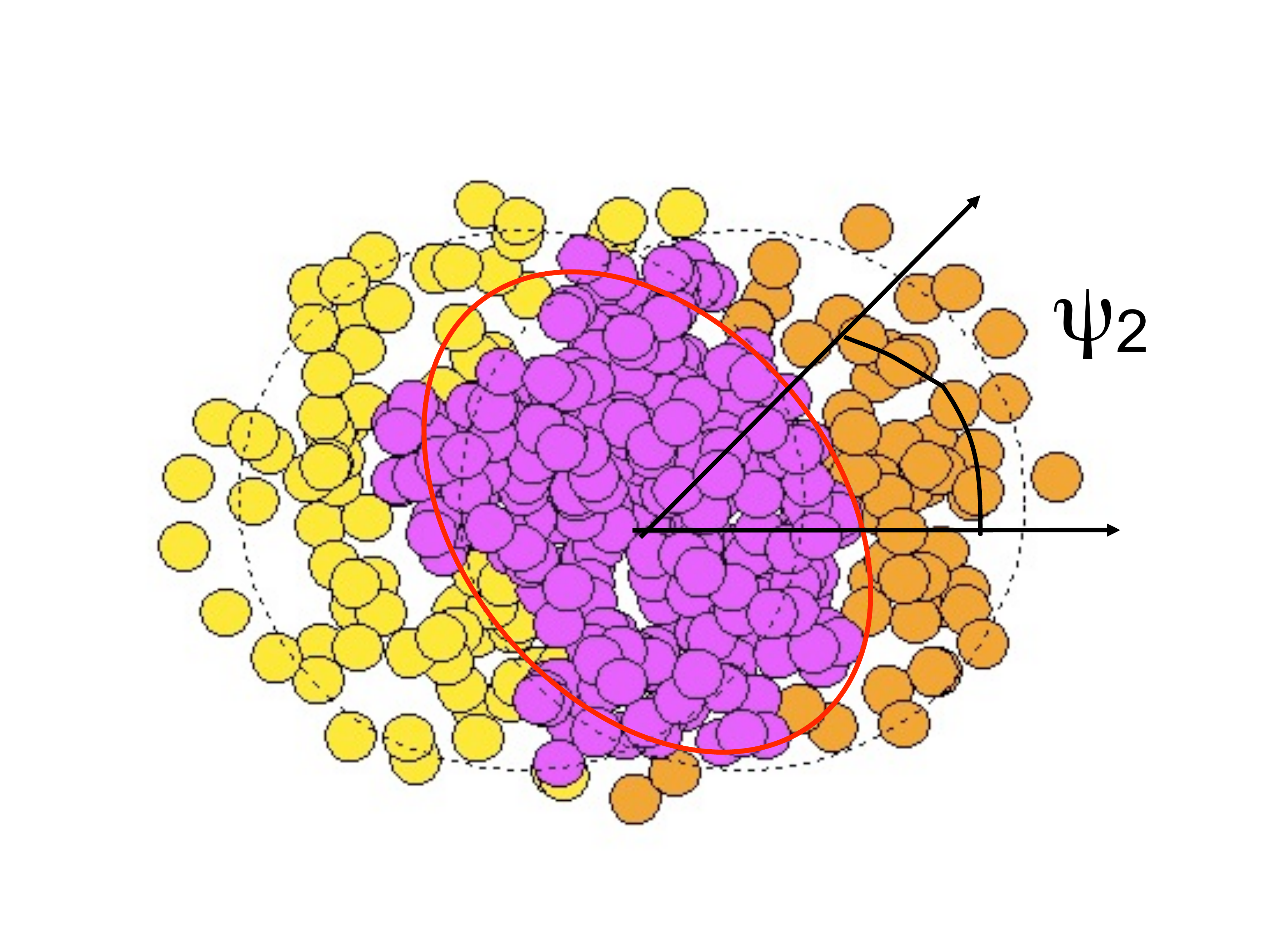}
\end{center}
\caption{\label{burakellipse}Interaction region projected on the plane perpendicular to the collision axis. Circles indicate the positions of nucleons within nuclei just before the collision. Purple nucleons are the participants, which undergo at least one collision. The positions of participants define an ellipse marked as a red line. $\Psi_2$ is the azimuthal angle of the minor axis, along which elliptic flow develops.}
\end{minipage}\hspace{2pc}%
\begin{minipage}{18pc}
\begin{center}
\includegraphics[width=13pc]{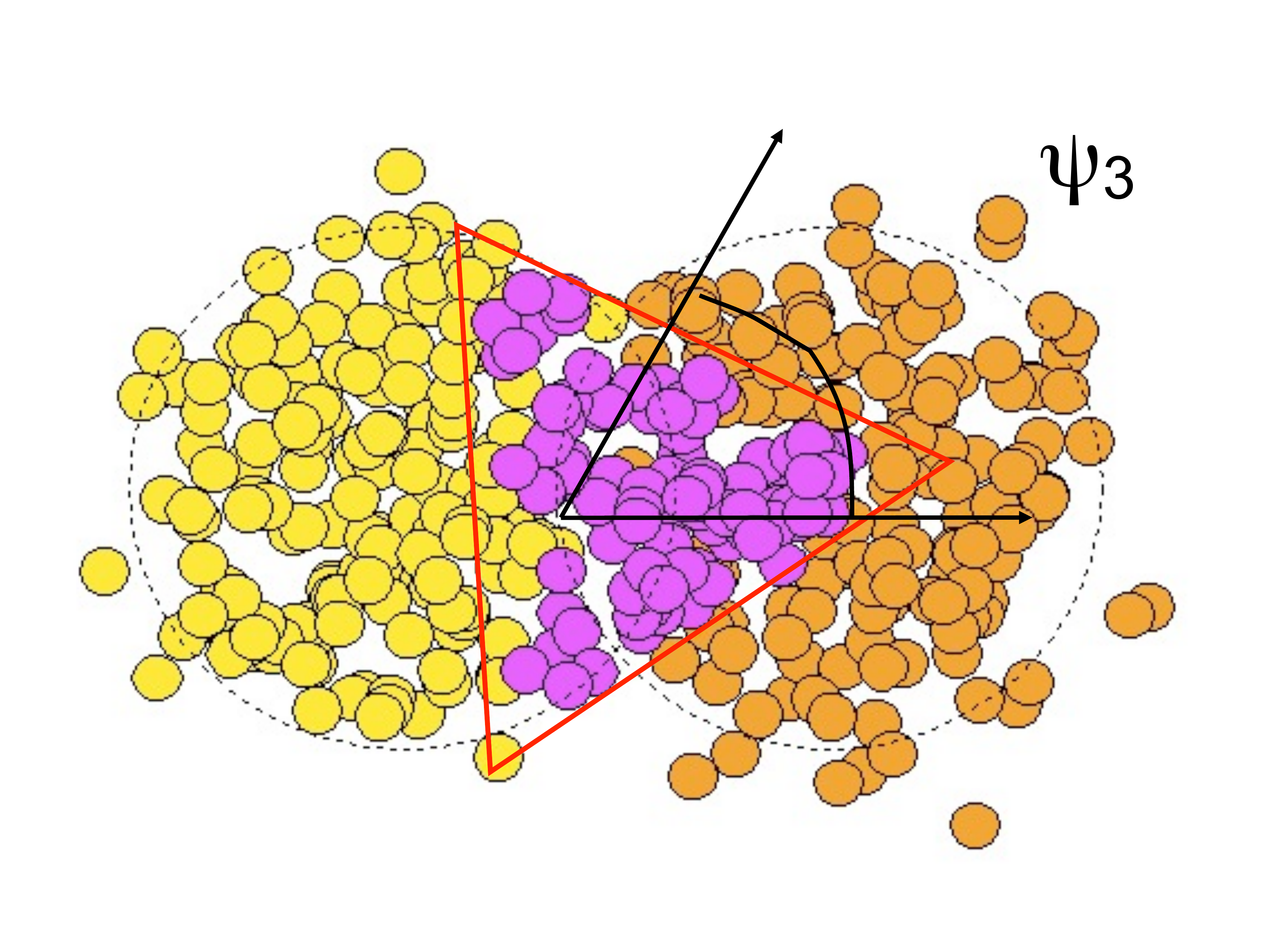}
\end{center}
\caption{\label{buraktriangle}(courtesy B. Alver) Another event, which has been chosen because the distribution of participant nucleons draws a triangle in the transverse plane. By analogy with 
figure~\ref{burakellipse}, one denotes by $\Psi_3$ the flat axis of
the triangle.
\phantom{coucou }
\phantom{coucou }
\phantom{coucou }
\phantom{coucou }
\phantom{coucou }
\phantom{coucou }
\phantom{coucou }
\phantom{coucou }
\phantom{coucou }
\phantom{coucou }
\phantom{coucou }
\phantom{coucou }
\phantom{coucou }
\phantom{coucou }
\phantom{coucou }
\phantom{coucou }
\phantom{coucou }
}
\end{minipage} 
\end{figure}
The breakthrough came a few months ago, when Alver and
Roland~\cite{Alver:2010gr} proposed such a mechanism. 
This mechanism involves fluctuations, which I have already 
mentioned, and now explain in more detail. Due to Lorentz 
contraction, the time scale of a heavy-ion collision at 
RHIC is so short that the positions of nucleons within nuclei are
frozen during the collision. The collision acts as a quantum
measurement process: the positions of nucleons are random numbers,
whose probability is determined by the wave function of the nucleus. 
Now, each nucleon-nucleon collision produces several particles whose
rapidity span a large interval. One therefore expects that at all
rapidities, the initial density profile has ``hot spots'' at points in
the transverse plane where there are more nucleons. 
These fluctuations are playing an increasingly important role in
heavy-ion phenomenology.
The new idea of Alver and Roland is that in some events, fluctuations
may create a triangular shape (see
figure~\ref{buraktriangle}). 
In the same way as elliptic flow develops along the minor axis of the ellipse, {\it triangular\/} flow develops along the flat side of the triangle. By analogy with equations~(\ref{otherdefv2}) and (\ref{v22}), we write 
\begin{equation}
\langle e^{3 i \phi}\rangle=v_3 e^{3 i \Psi_3}\ \ \ \ \ \ \ \langle\cos 3\Delta\phi\rangle=(v_3)^2.
\end{equation}
In the same way as $v_2$ is proportional to the initial eccentricity,
$v_3$ turns out to be proportional to the {\it triangularity\/},
$\varepsilon_3$~\cite{Alver:2010gr}, a dimensionless number which
characterizes how triangular the initial distribution is. 
This triangularity is produced by statistical fluctuations. 
For central collisions shown in figure~\ref{phobosAu}, however, the
eccentricity also results from statistical
fluctuations~\cite{Alver:2006wh}, and is of the same order of
magnitude. One therefore expects a similar magnitude for both 
$\cos(2\Delta\phi)$ and $\cos(3\Delta\phi)$ components. 
The triangular term is smaller because viscous damping is larger for
$v_3$ than for $v_2$~\cite{Alver:2010dn}. 
However, $v_3$ increases more rapidly with $p_t$ than $v_2$, and naturally
produces a shoulder structure of the correlation function 
(see figure~\ref{corr}) with a high-$p_t$ trigger, similar to the one observed 
experimentally~\cite{Adler:2005ee,Abelev:2008nda}.
Note that triangular flow appears naturally in hydrodynamic
simulations with fluctuating initial
geometries~\cite{Takahashi:2009na,Petersen:2010cw}. 

\section{Conclusions and perspectives}

There is now a wide consensus that
heavy-ion collisions at ultrarelativistic energies produce a fluid
with a small viscosity over entropy ratio. 
Relativistic viscous hydrodynamics describes the expansion of this
fluid, therefore it should explain the bulk of  
the observables: momentum spectra, anisotropies of identified
particles in the soft sector (below $p_t\sim 2-3$~GeV), as well as
Bose-Einstein correlations (not covered in this talk). 
The recent developments on triangular flow have shown that even the detailed
structure of correlations is naturally explained by flow. 
At LHC, the higher density will result in an increased lifetime for
the fluid, and heavy-ion collisions should be dominated by collective 
flow. On the theoretical side, viscous hydrodynamics is still in its
infancy, and much work remains to be done in order to properly
incorporate chemical~\cite{Monnai:2010qp} and thermal~\cite{Luzum:2010ad}
freeze-out. 

Our understanding of nucleus-nucleus collisions is hindered by our
poor knowledge of the initial state. In particular, quantum 
fluctuations, resulting from the fact that colliding nuclei are quantum
objects, are important for the little bang. In fact, the role of
fluctuations is in several respects the same as in big-bang
physics~\cite{Mishra:2007tw,Mishra:2008dm}. 
Present models of fluctuations are rather crude. 
There has been huge progress in our understanding of the 
wavefunction of a nucleus at high-energy, and this should be exploited 
fully to estimate quantities relevant to heavy-ion phenomenology, such
as the initial eccentricity~\cite{Lappi:2006xc} and fluctuations. 

\ack
I thank Fran\c cois Gelis and Kenji Fukushima for discussions, Burak
Alver, Ioannis Bouras and Matt Luzum, who provided some 
of the figures, and the Department of Theoretical physics of the 
Tata Institute of Fundamental Research, Mumbai, where this
contribution was written, for its kind hospitality.

\section*{References}
\bibliography{inpc}

\end{document}